\documentclass[aps,
prl,
floatfix,
superscriptaddress,
twocolumn,
footinbib,
longbibliography]{revtex4-2}
\usepackage{amssymb}
\usepackage{amsmath}
\usepackage{amsfonts}
\usepackage{appendix}
\usepackage{bm}
\usepackage[caption=false,subrefformat=parens]{subfig}
\usepackage{graphicx}
\usepackage{epsfig}
\usepackage{epstopdf}
\usepackage{balance}
\usepackage[dvipsnames]{xcolor}
\usepackage{calc}
\usepackage{natbib}
\usepackage[colorlinks,
linkcolor=blue,
anchorcolor=blue,
citecolor=blue,
urlcolor=blue]{hyperref}
\usepackage{lipsum}
\usepackage{braket}
\usepackage{subfig}
\usepackage{hyperref}
\usepackage{verbatim}
\usepackage{cleveref}

\newcommand{\revision}[1]{{\color{blue}{#1}}}
\begin{document}
\title{Atomic-scale observation of $d$-$\pi$-$d$ spin coupling in coordination structures}
\author{Xue Zhang}
\thanks{These authors contributed equally to this work.}
\affiliation{Center for Carbon-Based Electronics and Key Laboratory for the Physics and Chemistry of Nanodevices, School of Electronics, Peking University, Beijing 100871, China}
\affiliation{Spin-X Institute, School of Microelectronics, State Key Laboratory of Luminescent Materials and Devices, South China University of Technology, Guangzhou 511442, China}
\author{Xin Li}
\thanks{These authors contributed equally to this work.}
\affiliation{Center for Carbon-Based Electronics and Key Laboratory for the Physics and Chemistry of Nanodevices, School of Electronics, Peking University, Beijing 100871, China}
\author{Jie Li}
\thanks{These authors contributed equally to this work.}
\affiliation{Center for Carbon-Based Electronics and Key Laboratory for the Physics and Chemistry of Nanodevices, School of Electronics, Peking University, Beijing 100871, China}
\author{Haoyang Pan}
\thanks{These authors contributed equally to this work.}
\affiliation{Spin-X Institute, School of Chemistry and Chemical Engineering, South China University of Technology, Guangzhou 510641, China}
\author{Minghui Yu}
\affiliation{Center for Carbon-Based Electronics and Key Laboratory for the Physics and Chemistry of Nanodevices, School of Electronics, Peking University, Beijing 100871, China}
\author{Yajie Zhang}
\email{yjzhang11@pku.edu.cn}
\affiliation{Center for Carbon-Based Electronics and Key Laboratory for the Physics and Chemistry of Nanodevices, School of Electronics, Peking University, Beijing 100871, China}
\author{Gui-Lin Zhu}
\affiliation{School of Physics, Huazhong University of Science and Technology, Wuhan 430074, China}
\author{Zhen Xu}
\affiliation{Spin-X Institute, School of Chemistry and Chemical Engineering, South China University of Technology, Guangzhou 510641, China}
\author{Ziyong Shen}
\affiliation{Center for Carbon-Based Electronics and Key Laboratory for the Physics and Chemistry of Nanodevices, School of Electronics, Peking University, Beijing 100871, China}
\author{Shimin Hou}
\affiliation{Center for Carbon-Based Electronics and Key Laboratory for the Physics and Chemistry of Nanodevices, School of Electronics, Peking University, Beijing 100871, China}
\author{Yaping Zang}
\affiliation{BNLMS, Key Laboratory of Organic Solids, Institute of Chemistry, Chinese Academy of Sciences, Beijing 100190, China}
\author{Bingwu Wang}
\affiliation{BNLMS, Beijing Key Laboratory of Magnetoelectric Materials and Devices, College of Chemistry and Molecular Engineering, Peking University, Beijing 100871, China}
\author{Kai Wu}
\affiliation{BNLMS, College of Chemistry and Molecular Engineering, Peking University, Beijing 100871, China}
\author{Shang-Da Jiang}
\affiliation{Spin-X Institute, School of Chemistry and Chemical Engineering, State Key Laboratory of Luminescent Materials and Devices, Guangdong-Hong Kong-Macao Joint Laboratory of Optoelectronic and Magnetic Functional Materials, South China University of Technology, Guangzhou 511442, China}
\author{Ivano E. Castelli}
\affiliation{Department of Energy Conversion and Storage, Technical University of Denmark, DK-2800 Kongens Lyngby, Denmark}
\author{Lianmao Peng}
\affiliation{Center for Carbon-Based Electronics and Key Laboratory for the Physics and Chemistry of Nanodevices, School of Electronics, Peking University, Beijing 100871, China}
\author{Per Hedeg{\aa}rd}
\email{hedegard@nbi.ku.dk}
\affiliation{Niels Bohr Institute, University of Copenhagen, DK-2100 Copenhagen, Denmark}
\author{Song Gao}
\affiliation{Key Laboratory of Bioinorganic and Synthetic Chemistry of Ministry of Education, School of Chemistry, IGCME, GBRCE for Functional Molecular Engineering, Sun Yat-Sen University, Guangzhou 510275, China}
\affiliation{Spin-X Institute, School of Chemistry and Chemical Engineering, State Key Laboratory of Luminescent Materials and Devices, Guangdong-Hong Kong-Macao Joint Laboratory of Optoelectronic and Magnetic Functional Materials, South China University of Technology, Guangzhou 511442, China}
\affiliation{BNLMS, Beijing Key Laboratory of Magnetoelectric Materials and Devices, College of Chemistry and Molecular Engineering, Peking University, Beijing 100871, China}
\author{Jing-Tao L\"u}
\email{jtlu@hust.edu.cn}
\affiliation{School of Physics, Huazhong University of Science and Technology, Wuhan 430074, China}
\author{Yongfeng Wang}
\email{yongfengwang@pku.edu.cn}
\affiliation{Center for Carbon-Based Electronics and Key Laboratory for the Physics and Chemistry of Nanodevices, School of Electronics, Peking University, Beijing 100871, China}

\begin{abstract}
Spin coupling between magnetic metal atoms and organic radicals plays a pivotal role in high-performance magnetic materials. The complex interaction involving multi-spin centers in bulk materials makes it challenging to study spin coupling at the atomic scale. Here, we investigate the $d$-$\pi$-$d$ spin interaction in well-defined metal-organic coordinated structures composed of two iron (Fe) atoms and four all-trans retinoic acid (ReA) molecules, using low-temperature scanning tunneling microscopy and atomic force microscopy. The ReA molecule is turned into a spin-$1/2$ radical state by dehydrogenation, facilitating strong magnetic coupling with the coordinated Fe atoms. Comprehensive theoretical analysis, based on density functional theory and valence bond theory, further elucidates the intrinsic mechanism of ferrimagnetic spin coupling in the coordination structure. Specifically, simultaneous antiferromagnetic coupling of Fe dimer to ReA radicals parallelizes the dimer spin orientation. This work contributes to the fundamental understanding of spin interaction in metal-organic coordination structures and provides microscopic insights for designing advanced magnetic materials.
\end{abstract}

\maketitle

{\emph{Introduction}--}
%
Magnetic materials are the cornerstone of electronics, spintronics and quantum information science. Compared to the common inorganic counterparts, magnetic metal-organic materials have the advantages of lightweight and low-cost, 
whose structures and functions allow tuning through coordination between transition metal atoms and organic ligands\cite{Perlepe2020}. However, they suffer from weak super-exchange interaction among metal $d$ orbitals, leading to low magnetic ordering temperature. When the ligand is transformed to radical state, strong  direct exchange interaction between the metal $d$ orbital and the radical $\pi$ system may be used to enhance the long-range spin order through $d$-$\pi$-$d$ coupling, boosting the order temperature\cite{Perlepe2020,LiYang2023,ChengYang2024}. Bulk material contains multiple spin centers with complex spin interaction. If the basic metal-organic unit can be precisely constructed, and the $d$-$\pi$-$d$ spin interaction, as well as its response to chemical environment, can be quantitatively studied at the atomic scale, it can help to establish the relationship between microscopic spin interaction and macroscopic magnetism. 

Scanning tunneling microscope (STM) is an ideal tool to study atomic scale spin interaction, through measurement of the Kondo resonance and inelastic spin excitation spectrum\cite{LiJiutao1998,Madhavan1998,ZhaoAidi2005,Komeda2011,Mugarza2011,Franke2011,Minamitani2012,Zhang2013,Verlhac2019,Friedrich2024,Natterer2017,Trishi2023}. Different types of spin interaction among molecules and atoms have been investigated\cite{Hirjibehedin2006,Bork2011,Spinelli2015,Ternes2017,Choi2019,Loth2012,Ako2011,WangDimas2022,Mishra2020,SongLu2024,SunLorente2022,ZhaoWang2023,HeWang2022}. However, systematic study of $d$-$\pi$-$d$ spin interaction is still challenging due to at least two aspects. One is the precise construction of such system, and the other is the atomic precision measurement of spin interaction and quantitative understanding of its microscopic mechanism. 

In this work, a combination of surface coordination reaction and single-molecule spin manipulation is used to construct a well-defined $d$-$\pi$-$d$ system. Through on-demand manipulation of the molecular spin state, different types of spin interaction are precisely realized and quantitatively investigated with low-temperature STM, assisted by qPlus atomic force microscopy (AFM). Density functional theory (DFT) calculations and valence bond theory (VBT) elucidate the intrinsic mechanism of ferrimagnetic spin coupling of the $d$-$\pi$-$d$ structure in the $\big\uparrow$-$\downarrow$-$\big\uparrow$ form. Additionally, through STM manipulation, we also reveal how the spin interaction is affected by atomic scale change of the chemical environment. 

\begin{figure}
\includegraphics[width=\linewidth]{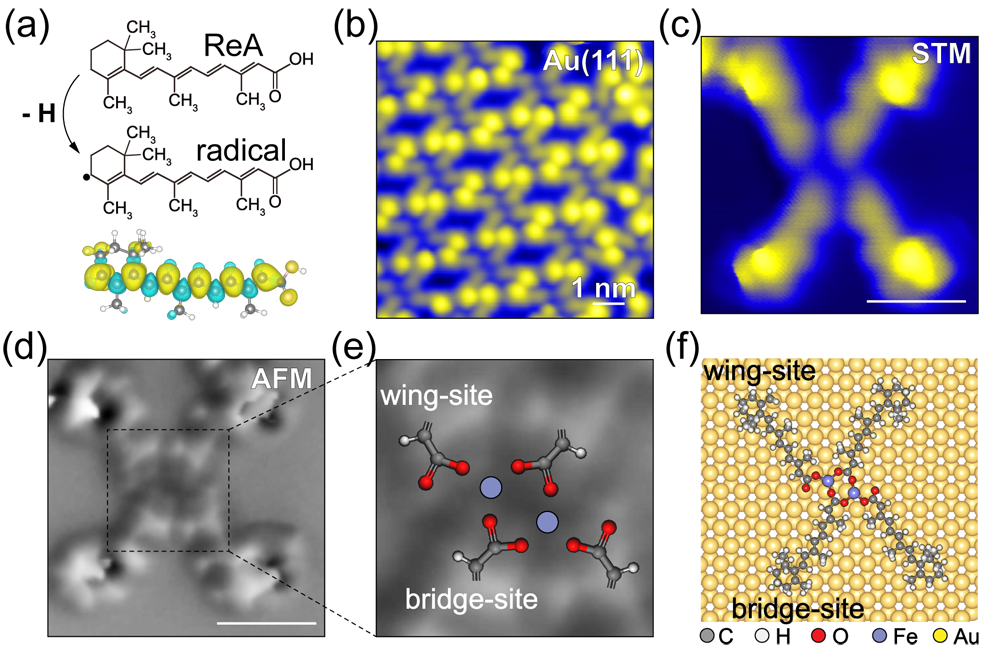}
\caption{{\bf Metal-organic coordinated structure formed by ReA tetramer and Fe dimer.} (a) A pristine ReA molecule can be transformed into a radical by dehydrogenation (up panel). Spin-density distribution of the transformed radical calculated from DFT (down panel). The yellow and blue color represents spin up and down, respectively. (b) STM image of Fe-ReA coordinated tetramers on Au(111). (c-d) STM (c) and AFM (d) images of one tetramer unit acquired using a Co-functionalized tip. (e) Zoom-in of the AFM image in (d) overlaid by proposed atomic structure. (f) DFT optimized tetramer structure on Au(111). Details of the calculation is in Sec. 1.1 of the Supplementary Materials (\revision{SM}). STM parameters: (b) Voltage bias $V_{\rm b} = 60$ mV, current $I_{\rm set} = 53$ pA; (c) $V_{\rm b} = -100$ mV, $I_{\rm set} = 20$ pA. AFM parameters: (d, e) feedback disabled at $V_{\rm b} = -100$ mV, $I_{\rm set} = 20$ pA on substrate with amplitude $A = 100$ pm and resonance frequency $f_0 = 22.4$ kHz. White scale bar: 1 nm.}
\label{fig:1}
\end{figure}

{\emph{Experimental and DFT Results}--}
The all-trans retinoic acid [ReA, Fig.~\ref{fig:1}(a)] molecules and Fe atoms form coordinated structures on Au(111) [Figs.~\ref{fig:1}(b-c)]. The molecular carboxyl groups are deprotonated during coordination with Fe\cite{Langner2007self,tait2008metal,vcechal2014convergent,dong2016self}. The methyl groups and oxygen atoms in the carboxyl group are clearly identifiable in the qPlus AFM images obtained with a CO tip [Figs.~\ref{fig:1}(d, e)]. 
%
Based on the AFM image, it is proposed that two Fe atoms are located at the center of the ReA tetramer and each Fe atom interacts with ReA via three-fold Fe-O coordination bonds [Fig.~\ref{fig:1}(e)], forming the basic structural unit. Two opposite ReA molecules (bridge-site, B) coordinate with Fe dimer through two oxygen, while the other two (wing-site, W) coordinate with only one Fe through one oxygen. Similar coordination structures have been previously reported\cite{Zhang2013,langner2012selective} and the DFT optimized structure [Fig.~\ref{fig:1}(f)] agrees well with the experimental topography. 

ReA molecule is a spin singlet but can transit into a radical with spin 1/2 via dehydrogenation ignited by applying an electrical pulse to the bulky 1,3,3-trimethylcyclohexene group\cite{Karan2016,Bocquet2019,HeWang2022} [Fig.~\ref{fig:1}(a)]. Single ReA radical features a Kondo resonance in the differential conductance ($dI/dV$) spectrum\cite{HeWang2022}. Our previous study on ReA radical pair in a self-assemble array has demonstrated splitting of the Kondo resonance due to their exchange coupling mediated by hydrogen bonds\cite{Karan2016}. 

\begin{figure}
		\includegraphics[width=\linewidth]{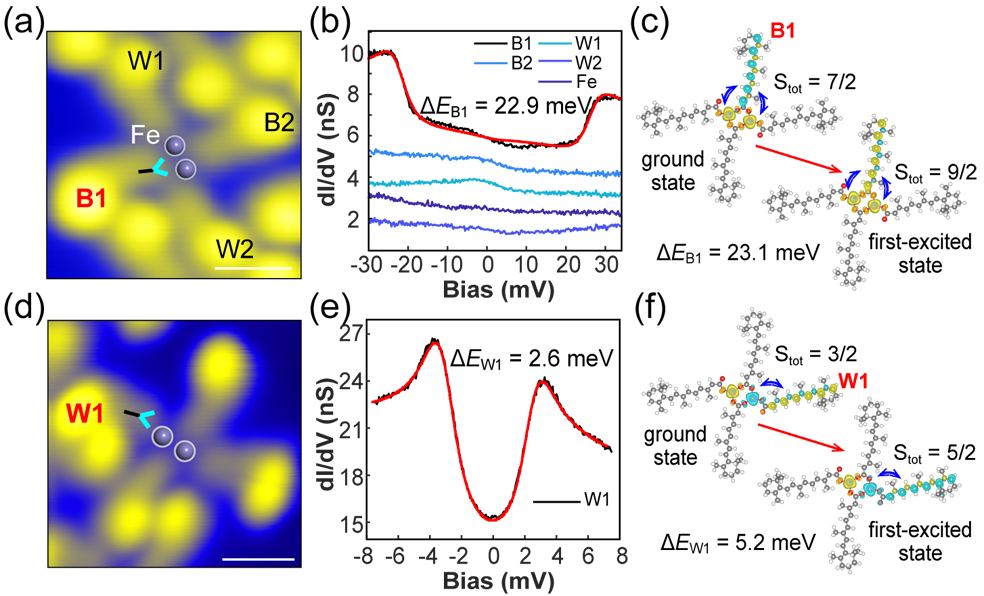}
\caption{{\bf Spin coupling between one ReA radical and Fe.} (a, d) STM images of the tetramer when a bridge-site (a) and a wing-site (d) ReA is turned into radical (red label), respectively. (b, e) The corresponding $dI/dV$ spectra measured on the labeled molecules or Fe in (a) and (c), respectively. The spectra of B2, W1, W2 and Fe in (b) have been shifted downward for 2, 3, 4 and 5 nS, respectively, for clarity. The red lines overlapped on the original $dI/dV$ spectra are theoretical fitting using perturbation theory (details in \revision{SM, Sec. 1.2})\cite{Ternes2015}. (c, f) Spin configurations of the ground and first-excited states and the corresponding spin excitation energy $\Delta E_{\rm B1}/\Delta E_{\rm W1}$. STM parameters: (a) $V_{\rm b} = 30$ mV, $I_{\rm set} = 43$ pA. (c) $V_{\rm b} = 30$ mV, $I_{\rm set} = 53$ pA. $dI/dV$ condition: (b) $V_{\rm b} = 30$ mV, $I_{\rm set} = 200$ pA; (d) $V_{\rm b} = -7.5$ mV, $I_{\rm set} = 200$ pA. The modulation voltage $V_{\rm mod}$ is 0.6 mV for $dI/dV$ measurement. The tip is positioned above the center of the 1,3,3-trimethylcyclohexene group during $dI/dV$ measurements throughout the paper unless claimed specifically.}
\label{fig:2}
\end{figure}

Here, we first transform a bridge-site ReA [B1, Fig.~\ref{fig:2}(a)] into a radical, which becomes brighter in the STM image (see also \revision{SM, Sec. 2.1}). The corresponding $dI/dV$ spectrum measured on the radical exhibits a step-like spin excitation feature, while the spectra at other positions are featureless [Fig.~\ref{fig:2}(b)].
Numerical fitting of the $dI/dV$ spectrum, based on the scattering theory (\revision{SM. Sec. 1.2})\cite{Appelbaum1967,Ternes2015}, yields a spin excitation energy of $\Delta E_{\rm B1} = 22.9$ meV for the radical-Fe coupling [Fig.~\ref{fig:2}(b)]. 
In contrast, when a wing-site ReA is turned into radical, a much smaller spin excitation energy $\Delta E_{\rm W1} = 2.6$ meV is observed [Fig.~\ref{fig:2}(d, e)]. Generally, the excitation energy measured on a bridge-site radical ranges from 19 to 25 meV, whereas it varies from 0.1 to 3 meV on a wing site (\revision{Sec 2.2 of SM}). We attribute this variation to the differences in molecular adsorption configuration, which may influence the metal-organic spin interaction and its competition with molecule-substrate spin interaction (discussed below). Additionally, $dI/dV$ spectra measured at different positions on the radical changes due to its delocalized spin density distribution (\revision{SM, Sec. 2.3}). 

DFT calculations excluding metal substrate agree well with the above experimental findings. 
Fe atom and the radical are at spin $2$ and $1/2$ state, respectively. Moreover, at the ground state the radical spin prefers to couple to the coordinated Fe spin antiferromagnetically in both bridge and wing cases [Fig.~\ref{fig:2}(c, f)], while the first-excited state corresponds to a ferromagnetic coupling between them.
For coupling to a bridge-site, the ground and first-excited state has a total spin (two Fe atoms plus one radical) of 7/2 and 9/2, respectively. This results in an excitation energy of $\Delta E_{\rm B1} = 23.1$ meV [Fig.~\ref{fig:2}(c)], in excellent agreement with the experimental results [Fig.~\ref{fig:2}(b)]. For coupling to a wing-site, total spin (one Fe atom and one radical) of 3/2 and 5/2 for the ground and first-excited state is obtained, respectively, resulting in an excitation energy of $\Delta E_{\rm W1}=5.2$ meV [Fig.~\ref{fig:2}(f)], larger than the experimental value between 0.1-3 meV. This is also attributed to the effect of metal substrate and 
further verified by subsequent experimental characterization discussed later. 

\begin{figure}
\includegraphics[width=\linewidth]{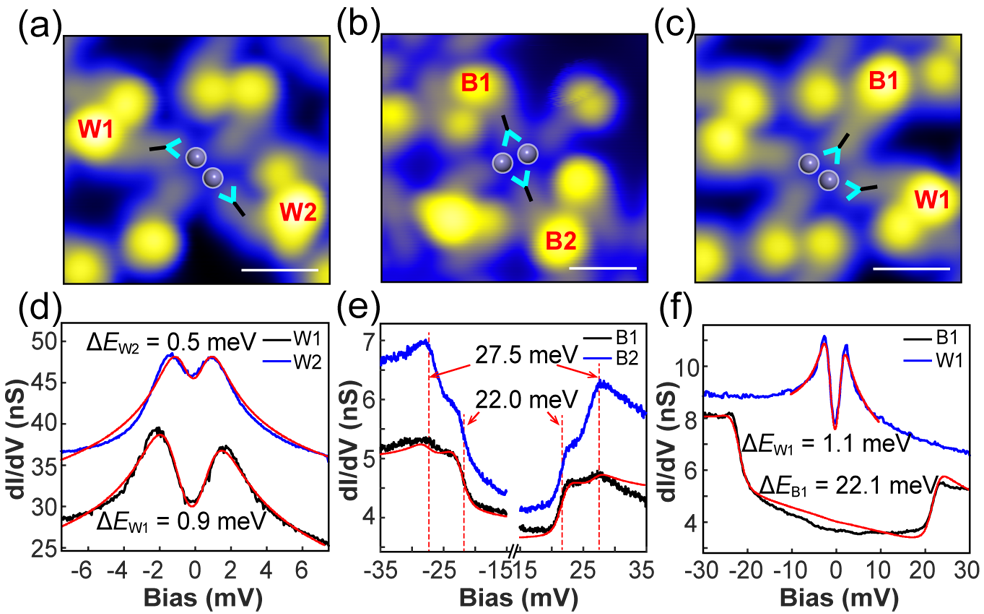}
\caption{{\bf Spin coupling between two ReA radicals and Fe.} (a-c) STM images showing different cases where two ReA  of different coordination sites (shown in red) are in radical states: (a) wing-wing-site (W1, W2), (b) bridge-bridge-site (B1, B2), (c) bridge-wing-site (B1, W1). Colored tridents highlight the carboxyl groups. (d-f) $dI/dV$ spectra together with the excitation energy measured on the ReA radicals marked in (a-c). The red lines are theoretical fitting as in Fig.~(\ref{fig:2}).   STM parameters: (a) $V_{\rm b} = 30$ mV, $I_{\rm set} = 83$ pA; (b, c) $V_{\rm b} = 30$ mV, $I_{\rm set} = 53$ pA. $dI/dV$ condition: (d) $V_{\rm b} = 30$ mV, $I_{\rm set} = 180$ pA; (e) $V_{\rm b} = 7.5$ mV, $I_{\rm set} = 180$ pA; (f) $V_{\rm b} = -35$ mV (negative side) or 35 mV (positive side), $I_{\rm set} = 180$ pA. The $dI/dV$ spectrum of W1 in (d), W2 in (e) and B2 in (f) is shifted vertically by 0.5, 10, 0.5 nS, respectively.}
\label{fig:3}
\end{figure}

We proceed to investigate the spin coupling in the presence of two ReA radicals in the coordinated structure. Considering the two distinct coordination sites of ReA molecules, we examine three different combinations [Fig.~\ref{fig:3}(a–c)]. Firstly, when two wing-site molecules are transformed to radicals [Fig.~\ref{fig:3}(a)], we observe gapped $dI/dV$ spectra on both molecules [Fig.~\ref{fig:3}(d)] and each of them is very similar to that of the single wing-site radical case [Fig.~\ref{fig:2}(e)]. Secondly, when two bridge-site molecules are simultaneously turned into radicals [Fig.~\ref{fig:3}(b)], we observe clear double steps in the $dI/dV$ spectra, as highlighted by the dashed lines in Fig.~\ref{fig:3}(e). They correspond to two spin excitation energies, $\Delta E_1 = 22.0$ meV and $\Delta E_2 = 27.5$ meV with a ratio of $\Delta E_2/\Delta E_1=1.25 $. Thirdly, when one bridge-site and one wing-site are turned into radical states [Fig.~\ref{fig:3}(c)]. The $dI/dV$ spectrum measured on either radical [Fig.~\ref{fig:3}(f)] closely resembles that observed when only the corresponding single ReA molecule becomes radical. 
We further transformed all the four ReA molecules in a tetramer to radicals. The resulting spectra at the bridge sites resemble the case when only two bridge molecules are in radical states, similarly for the two wing sites (\revision{SM, Sec. 2.3}). These results suggest strong Fe dimer and inter-radical coupling only when the two bridge sites are in radical states. 

\begin{figure}
		\includegraphics[width=0.8\linewidth]{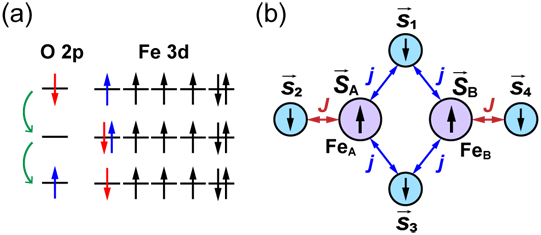}
\caption{{\bf Schematic model illustrating the spin coupling mechanism between ReA radicals and Fe atoms.} (a) Representative exchange process showing an unpaired electron from the oxygen $2p$ orbital swapping with an unpaired electron from the Fe $3d$ orbital, demonstrating antiferromagnetic coupling. (b) Four ReA radical spins (labeled as $1$ to $4$) coupling with two Fe spins ($A$ and $B$). Each ReA radical spin couples to Fe through the carboxyl group with exchange energy $j$ (bridge) and $J$ (wing), respectively. }
\label{fig:4}
\end{figure}

{\emph{Theoretical analysis}--}
Considering the delocalized spin distributed on the $\pi$ orbital, spin located at oxygen atoms can couple with Fe. A typical exchange process according to the VBT is illustrated in Fig.~\ref{fig:4}(a), which results in antiferromagnetic coupling. Conversely, electron exchange is forbidden if the coupling is ferromagnetic. Thus, we deduce that the ground state corresponds to the antiferromagnetic coupling between the radical and Fe. 
%
Accordingly, a Heisenberg spin Hamiltonian for the coordinated tetramer [Fig.~\ref{fig:4}(b)] can be used to understand the experimental results
\begin{align}
    H=j\vec{s}_{13}\cdot \vec{S}_{AB}+J\vec{s}_2\cdot\vec{S}_A+J\vec{s}_4\cdot \vec{S}_B.
    \label{eq:heisenberg}
\end{align}
Here, $\vec{s}_{13}=\vec{s}_1+\vec{s}_3$, with $\vec{s}_n$ ($n = 1-4$) the spin operator of the $n$-th radical labeled in Fig.~\ref{fig:4}(a), $\vec{S}_{AB}=\vec{S}_{A}+\vec{S}_{B}$ with $\vec{S}_A$ and $\vec{S}_B$ spin operators of the two Fe ions, and the positive $J$ ($j$) represents exchange coupling between a wing (bridge) site ReA radical and the coordinated Fe. Due to distinct coordination configuration, the bridge and wing sites have different coupling parameters with Fe. 

We start by considering the cases with only one ReA in radical state. 
In the case of a bridge radical coupling to two Fe ions, Eq.~(\ref{eq:heisenberg}) simplifies to $H_b=j\vec{s}_1\cdot\vec{S}_{AB}=j(\vec{S}_{\rm tot}^2-\vec{s}_1^2-\vec{S}_{AB}^2)/2$, with $\vec{S}_{\rm tot}=\vec{s}_1+\vec{S}_{AB}$. The ground state corresponds to a total spin ${S}_{\rm tot}=7/2$ and energy $E_1=-5j/2$, with $\vec{s}_1$ being antiparallel to both $\vec{S}_A$ and $\vec{S}_B$. For the first excited state, all spins are parallel with ${S}_{\rm tot}=9/2$ and energy $E_2=2j$. The excitation energy is thus $\Delta E_{\rm B1}=|E_1-E_2|=9j/2$. For a wing-site ReA radical coupling to one Fe atom, the Hamiltonian is $H_w=J\vec{s}_2\cdot\vec{S}_A$. Similar analysis gives the ground state with ${S}_{\rm tot}=3/2$, $E_1=-3J/2$ and the first excited state ${S}_{\rm tot}=5/2$, $E_2=J$. The corresponding excitation energy is $\Delta E_{\rm W1}=5J/2$. In both cases, the excitation corresponds physically to flip of radical spin. 
From the experimentally measured spectra in Fig.~\ref{fig:2}(b, e), the exchange coupling parameter $j=5.1$ meV and  $J=1.0$ meV can be deduced.

We now consider the cases where two radical spins are involved, and  
focus on the most interesting situation when the two bridge-site molecules are in radical states. Equation~(\ref{eq:heisenberg}) simplifies to  
$H_{bb}=j\vec{s}_{13}\cdot\vec{S}_{AB}$.
The ground state is obtained when $\vec{s}_{13}$ is parallel or anti-parallel to $\vec{S}_{AB}$. The length of $\vec{s}_{13}$ takes 1 or 0, while the maximum length of $\vec{S}_{AB}$ is 4. Thus, $S_{\rm tot}$ can have values $4+1=5$, $4$, $4$ and $4-1=3$. The state with $S_{\rm tot} = 5$ represents the ferromagnetic coupling between the radical and Fe spin. Notably, there are two states with ${S}_{\rm tot} = 4$. One corresponds to ${s}_{13}=1$ but the $z$ component equals to 0, while the other represents $s_{13}=0$. The state with ${S}_{\rm tot} = 3$ represents antiferromagnetic coupling between radical and Fe. Writing the Hamiltonian in the following form
\begin{align}
    H_{bb} &=\frac{j}{2}\left[S_{\rm tot}(S_{\rm tot}+1)-s_{13}(s_{13}+1)-S_{AB}(S_{AB}+1)\right], 
\end{align}
the energies of these four states are obtained as $4j$, $-j$, $0$ and $-5j$,  respectively. Since $j>0$, the first and second excitation energy are  $\Delta E_1=4j$ and $\Delta E_2=5j$, respectively, with a theoretical ratio $\Delta E_2/\Delta E_1=1.25$. This is in perfect agreement with the experimentally measured values of $\Delta E_1=22.0$ meV and $\Delta E_2=27.5$ meV with a ratio of 1.25 [Fig.~\ref{fig:3}(e)], verifying the presence of magnetic coupling between the two bridge-site radical spins through their mutually coordinated Fe atoms.  

The rest two cases in Fig.~\ref{fig:3}(a, d) and \ref{fig:3}(c, f) involve at least one radical at wing site. In the former case, since we have ignored the direct exchange coupling between two Fe atoms, the two Fe-radical pair decouple. In the latter case, the large difference between $J$ and $j$ renders their coupling negligible. 
Consequently, both of them show negligible coupling between the two radical spins in the spectra. 
%
%

\begin{figure}
\includegraphics[width=\linewidth]{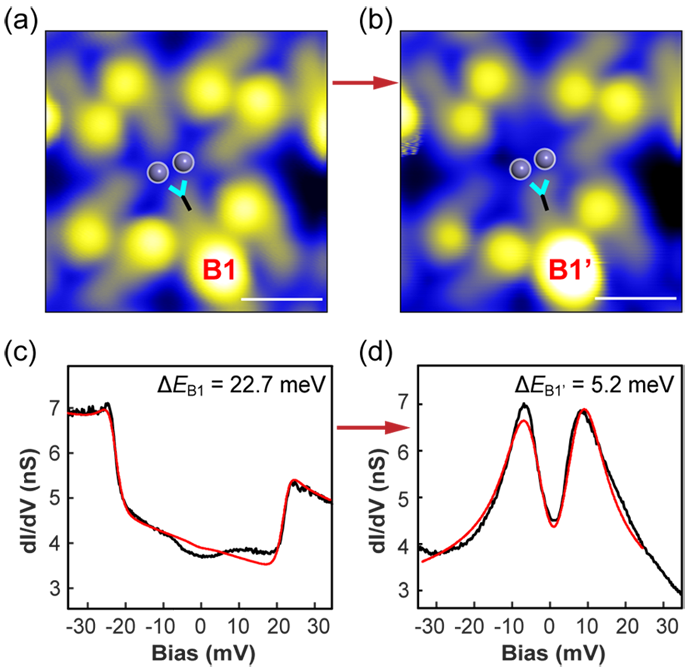}
\caption{{\bf Competition between radical-Fe coupling and molecule-substrate interaction.} (a, b) STM images of a bridge-site ReA radical coupling to Fe atoms before (marked as B1) and after (marked as B1') applying electrical pulses. (c, d) $dI/dV$ spectra measured on the ReA radical in (a) and (b), respectively. STM condition: constant-current mode, (a) $V_{\rm b} = 30$ mV, $I_{\rm set} = 43$ pA; (b) $V_{\rm b} = 30$ mV, $I_{\rm set} = 53$ pA; $dI/dV$ condition: $V_{\rm b} = -35$ mV, $I_{\rm set} = 250$ pA.}
\label{fig:5}
\end{figure}

{\emph{Effect of substrate}--}
Since both molecules and Fe atoms are on the metal substrate, there exists competition between the radical-Fe spin coupling and the radical spin coupling with itinerant substrate electrons. The latter can vary in strength, likely influenced by molecular conformational change or redistribution of the substrate density of states. This variability has led to a range of observed Kondo temperatures of ReA radical spanning from 4 K to 20 K\cite{Karan2016}. Here, we physically perturb the ReA radical using the STM tip and measure the resulting changes in the spectrum. Figure \ref{fig:5}(a) depicts a bridge-site radical exhibiting broad spin excitation steps [Fig.~\ref{fig:5}(c)], attributed to significant $d$-$\pi$ magnetic coupling. After the manipulation, the molecule appears brighter [Fig.~\ref{fig:5}(b)], and a pronounced Kondo feature emerge in the $dI/dV$ spectrum [Fig.~\ref{fig:5}(d)]. This indicates that the interaction between the radical spin and the metal substrate becomes dominant. Consequently, the spin excitation energy decreases substantially from 22.7 to 5.2 meV. 
Deviation of the larger calculated spin excitation energy from the experimentally observed values (Fig.~\ref{fig:2}) can be attributed to these complex interactions, as the DFT calculation of excitation energy does not account for the effect of substrate. 
This result highlights the importance of substrate-mediated interactions in determining the magnetic properties of surface-adsorbed molecular systems. 

{\emph{Conclusions}--}
In summary, we have investigated $d$-$\pi$-$d$ ferrimagnetic spin interaction in a coordinated metal-organic system formed by carboxylic molecular radicals and Fe atoms on surface. Differential conductance measurements confirm the spin coupling, revealing distinct spin excitation energies depending on two different coordination sites. Theoretical calculations combined with valence bond theory and Heisenberg model analysis show that the radical spins prefer to couple antiferromagnetically to Fe spins at both coordination sites. Our study provides valuable insights into achieving complex magnetic interactions based on $d$-$\pi$ coupling in coordinated metal-organic structures, with significant potential for developing advanced molecule-based magnetic materials and nanoscale spintronic devices.

{\emph{Acknowledgements}--}
This work was supported by National Natural Science Foundation of China (Grants No. 22225202, 92356309, 22132007, 21991132, 22202074, 22172002, 22273029) and Science and Technology Project of Guangzhou (grant agreement No. 2023A04J0671).
DFT calculations are carried out on TianHe-1A at National Supercomputer Center in Tianjin and supported by High-performance Computing Platform of Peking University. Experiments are supported by Peking Nanofab.

\bibliography{ref.bib}

\begin{thebibliography}{37}%
\makeatletter
\providecommand \@ifxundefined [1]{%
 \@ifx{#1\undefined}
}%
\providecommand \@ifnum [1]{%
 \ifnum #1\expandafter \@firstoftwo
 \else \expandafter \@secondoftwo
 \fi
}%
\providecommand \@ifx [1]{%
 \ifx #1\expandafter \@firstoftwo
 \else \expandafter \@secondoftwo
 \fi
}%
\providecommand \natexlab [1]{#1}%
\providecommand \enquote  [1]{``#1''}%
\providecommand \bibnamefont  [1]{#1}%
\providecommand \bibfnamefont [1]{#1}%
\providecommand \citenamefont [1]{#1}%
\providecommand \href@noop [0]{\@secondoftwo}%
\providecommand \href [0]{\begingroup \@sanitize@url \@href}%
\providecommand \@href[1]{\@@startlink{#1}\@@href}%
\providecommand \@@href[1]{\endgroup#1\@@endlink}%
\providecommand \@sanitize@url [0]{\catcode `\\12\catcode `\$12\catcode `\&12\catcode `\#12\catcode `\^12\catcode `\_12\catcode `\%12\relax}%
\providecommand \@@startlink[1]{}%
\providecommand \@@endlink[0]{}%
\providecommand \url  [0]{\begingroup\@sanitize@url \@url }%
\providecommand \@url [1]{\endgroup\@href {#1}{\urlprefix }}%
\providecommand \urlprefix  [0]{URL }%
\providecommand \Eprint [0]{\href }%
\providecommand \doibase [0]{https://doi.org/}%
\providecommand \selectlanguage [0]{\@gobble}%
\providecommand \bibinfo  [0]{\@secondoftwo}%
\providecommand \bibfield  [0]{\@secondoftwo}%
\providecommand \translation [1]{[#1]}%
\providecommand \BibitemOpen [0]{}%
\providecommand \bibitemStop [0]{}%
\providecommand \bibitemNoStop [0]{.\EOS\space}%
\providecommand \EOS [0]{\spacefactor3000\relax}%
\providecommand \BibitemShut  [1]{\csname bibitem#1\endcsname}%
\let\auto@bib@innerbib\@empty
\bibitem [{\citenamefont {Perlepe}\ \emph {et~al.}(2020)\citenamefont {Perlepe}, \citenamefont {Oyarzabal}, \citenamefont {Mailman}, \citenamefont {Yquel}, \citenamefont {Platunov}, \citenamefont {Dovgaliuk}, \citenamefont {Rouzieres}, \citenamefont {Negrier}, \citenamefont {Mondieig}, \citenamefont {Suturina}, \citenamefont {Dourges}, \citenamefont {Bonhommeau}, \citenamefont {Musgrave}, \citenamefont {Pedersen}, \citenamefont {Chernyshov}, \citenamefont {Wilhelm}, \citenamefont {Rogalev}, \citenamefont {Mathoniere},\ and\ \citenamefont {Clerac}}]{Perlepe2020}%
  \BibitemOpen
  \bibfield  {author} {\bibinfo {author} {\bibfnamefont {P.}~\bibnamefont {Perlepe}}, \bibinfo {author} {\bibfnamefont {I.}~\bibnamefont {Oyarzabal}}, \bibinfo {author} {\bibfnamefont {A.}~\bibnamefont {Mailman}}, \bibinfo {author} {\bibfnamefont {M.}~\bibnamefont {Yquel}}, \bibinfo {author} {\bibfnamefont {M.}~\bibnamefont {Platunov}}, \bibinfo {author} {\bibfnamefont {I.}~\bibnamefont {Dovgaliuk}}, \bibinfo {author} {\bibfnamefont {M.}~\bibnamefont {Rouzieres}}, \bibinfo {author} {\bibfnamefont {P.}~\bibnamefont {Negrier}}, \bibinfo {author} {\bibfnamefont {D.}~\bibnamefont {Mondieig}}, \bibinfo {author} {\bibfnamefont {E.~A.}\ \bibnamefont {Suturina}}, \bibinfo {author} {\bibfnamefont {M.-A.}\ \bibnamefont {Dourges}}, \bibinfo {author} {\bibfnamefont {S.}~\bibnamefont {Bonhommeau}}, \bibinfo {author} {\bibfnamefont {R.~A.}\ \bibnamefont {Musgrave}}, \bibinfo {author} {\bibfnamefont {K.~S.}\ \bibnamefont {Pedersen}}, \bibinfo {author} {\bibfnamefont {D.}~\bibnamefont {Chernyshov}}, \bibinfo {author}
  {\bibfnamefont {F.}~\bibnamefont {Wilhelm}}, \bibinfo {author} {\bibfnamefont {A.}~\bibnamefont {Rogalev}}, \bibinfo {author} {\bibfnamefont {C.}~\bibnamefont {Mathoniere}},\ and\ \bibinfo {author} {\bibfnamefont {R.}~\bibnamefont {Clerac}},\ }\bibfield  {title} {\bibinfo {title} {Metal-organic magnets with large coercivity and ordering temperatures up to 242$^\circ${C}},\ }\href {https://doi.org/10.1126/science.abb3861} {\bibfield  {journal} {\bibinfo  {journal} {Science}\ }\textbf {\bibinfo {volume} {370}},\ \bibinfo {pages} {587} (\bibinfo {year} {2020})}\BibitemShut {NoStop}%
\bibitem [{\citenamefont {Li}\ \emph {et~al.}(2023)\citenamefont {Li}, \citenamefont {Liu}, \citenamefont {Tang}, \citenamefont {Li}, \citenamefont {Ding}, \citenamefont {Liu}, \citenamefont {Fu}, \citenamefont {Dong}, \citenamefont {Li},\ and\ \citenamefont {Yang}}]{LiYang2023}%
  \BibitemOpen
  \bibfield  {author} {\bibinfo {author} {\bibfnamefont {X.}~\bibnamefont {Li}}, \bibinfo {author} {\bibfnamefont {Q.-B.}\ \bibnamefont {Liu}}, \bibinfo {author} {\bibfnamefont {Y.}~\bibnamefont {Tang}}, \bibinfo {author} {\bibfnamefont {W.}~\bibnamefont {Li}}, \bibinfo {author} {\bibfnamefont {N.}~\bibnamefont {Ding}}, \bibinfo {author} {\bibfnamefont {Z.}~\bibnamefont {Liu}}, \bibinfo {author} {\bibfnamefont {H.-H.}\ \bibnamefont {Fu}}, \bibinfo {author} {\bibfnamefont {S.}~\bibnamefont {Dong}}, \bibinfo {author} {\bibfnamefont {X.}~\bibnamefont {Li}},\ and\ \bibinfo {author} {\bibfnamefont {J.}~\bibnamefont {Yang}},\ }\bibfield  {title} {\bibinfo {title} {Quintuple function integration in two-dimensional cr(ii) five- membered heterocyclic metal organic frameworks via tuning ligand spin and lattice symmetry},\ }\href {https://doi.org/10.1021/jacs.2c12780} {\bibfield  {journal} {\bibinfo  {journal} {J. Am. Chem. Soc.}\ }\textbf {\bibinfo {volume} {145}},\ \bibinfo {pages} {7869} (\bibinfo {year}
  {2023})}\BibitemShut {NoStop}%
\bibitem [{\citenamefont {Cheng}\ \emph {et~al.}(2024)\citenamefont {Cheng}, \citenamefont {Li},\ and\ \citenamefont {Yang}}]{ChengYang2024}%
  \BibitemOpen
  \bibfield  {author} {\bibinfo {author} {\bibfnamefont {J.}~\bibnamefont {Cheng}}, \bibinfo {author} {\bibfnamefont {X.}~\bibnamefont {Li}},\ and\ \bibinfo {author} {\bibfnamefont {J.}~\bibnamefont {Yang}},\ }\bibfield  {title} {\bibinfo {title} {Room-temperature ferrimagnetism and size-modulated electronic structures in two-dimensional cluster-based metal-organic frameworks},\ }\href {https://doi.org/10.1007/s11426-023-1936-9} {\bibfield  {journal} {\bibinfo  {journal} {Sci. China Chem.}\ }\textbf {\bibinfo {volume} {67}},\ \bibinfo {pages} {1334} (\bibinfo {year} {2024})}\BibitemShut {NoStop}%
\bibitem [{\citenamefont {Li}\ \emph {et~al.}(1998)\citenamefont {Li}, \citenamefont {Schneider}, \citenamefont {Berndt},\ and\ \citenamefont {Delley}}]{LiJiutao1998}%
  \BibitemOpen
  \bibfield  {author} {\bibinfo {author} {\bibfnamefont {J.}~\bibnamefont {Li}}, \bibinfo {author} {\bibfnamefont {W.-D.}\ \bibnamefont {Schneider}}, \bibinfo {author} {\bibfnamefont {R.}~\bibnamefont {Berndt}},\ and\ \bibinfo {author} {\bibfnamefont {B.}~\bibnamefont {Delley}},\ }\bibfield  {title} {\bibinfo {title} {{Kondo} scattering observed at a single magnetic impurity},\ }\href {https://doi.org/10.1103/PhysRevLett.80.2893} {\bibfield  {journal} {\bibinfo  {journal} {Phys. Rev. Lett.}\ }\textbf {\bibinfo {volume} {80}},\ \bibinfo {pages} {2893} (\bibinfo {year} {1998})}\BibitemShut {NoStop}%
\bibitem [{\citenamefont {Madhavan}\ \emph {et~al.}(1998)\citenamefont {Madhavan}, \citenamefont {Chen}, \citenamefont {Jamneala}, \citenamefont {Crommie~M.},\ and\ \citenamefont {Wingreen~N.}}]{Madhavan1998}%
  \BibitemOpen
  \bibfield  {author} {\bibinfo {author} {\bibfnamefont {V.}~\bibnamefont {Madhavan}}, \bibinfo {author} {\bibfnamefont {W.}~\bibnamefont {Chen}}, \bibinfo {author} {\bibfnamefont {T.}~\bibnamefont {Jamneala}}, \bibinfo {author} {\bibfnamefont {F.}~\bibnamefont {Crommie~M.}},\ and\ \bibinfo {author} {\bibfnamefont {S.}~\bibnamefont {Wingreen~N.}},\ }\bibfield  {title} {\bibinfo {title} {Tunneling into a single magnetic atom: Spectroscopic evidence of the {Kondo} resonance},\ }\href {https://doi.org/10.1126/science.280.5363.567} {\bibfield  {journal} {\bibinfo  {journal} {Science}\ }\textbf {\bibinfo {volume} {280}},\ \bibinfo {pages} {567} (\bibinfo {year} {1998})}\BibitemShut {NoStop}%
\bibitem [{\citenamefont {Zhao}\ \emph {et~al.}(2005)\citenamefont {Zhao}, \citenamefont {Li}, \citenamefont {Chen}, \citenamefont {Xiang}, \citenamefont {Wang}, \citenamefont {Pan}, \citenamefont {Wang}, \citenamefont {Xiao}, \citenamefont {Yang}, \citenamefont {Hou},\ and\ \citenamefont {Zhu}}]{ZhaoAidi2005}%
  \BibitemOpen
  \bibfield  {author} {\bibinfo {author} {\bibfnamefont {A.}~\bibnamefont {Zhao}}, \bibinfo {author} {\bibfnamefont {Q.}~\bibnamefont {Li}}, \bibinfo {author} {\bibfnamefont {L.}~\bibnamefont {Chen}}, \bibinfo {author} {\bibfnamefont {H.}~\bibnamefont {Xiang}}, \bibinfo {author} {\bibfnamefont {W.}~\bibnamefont {Wang}}, \bibinfo {author} {\bibfnamefont {S.}~\bibnamefont {Pan}}, \bibinfo {author} {\bibfnamefont {B.}~\bibnamefont {Wang}}, \bibinfo {author} {\bibfnamefont {X.}~\bibnamefont {Xiao}}, \bibinfo {author} {\bibfnamefont {J.}~\bibnamefont {Yang}}, \bibinfo {author} {\bibfnamefont {J.~G.}\ \bibnamefont {Hou}},\ and\ \bibinfo {author} {\bibfnamefont {Q.}~\bibnamefont {Zhu}},\ }\bibfield  {title} {\bibinfo {title} {Controlling the {Kondo} effect of an adsorbed magnetic ion through its chemical bonding},\ }\href {https://doi.org/10.1126/science.111344} {\bibfield  {journal} {\bibinfo  {journal} {Science}\ }\textbf {\bibinfo {volume} {309}},\ \bibinfo {pages} {1542} (\bibinfo {year} {2005})}\BibitemShut
  {NoStop}%
\bibitem [{\citenamefont {Komeda}\ \emph {et~al.}(2011)\citenamefont {Komeda}, \citenamefont {Isshiki}, \citenamefont {Liu}, \citenamefont {Zhang}, \citenamefont {Lorente}, \citenamefont {Katoh}, \citenamefont {Breedlove},\ and\ \citenamefont {Yamashita}}]{Komeda2011}%
  \BibitemOpen
  \bibfield  {author} {\bibinfo {author} {\bibfnamefont {T.}~\bibnamefont {Komeda}}, \bibinfo {author} {\bibfnamefont {H.}~\bibnamefont {Isshiki}}, \bibinfo {author} {\bibfnamefont {J.}~\bibnamefont {Liu}}, \bibinfo {author} {\bibfnamefont {Y.-F.}\ \bibnamefont {Zhang}}, \bibinfo {author} {\bibfnamefont {N.}~\bibnamefont {Lorente}}, \bibinfo {author} {\bibfnamefont {K.}~\bibnamefont {Katoh}}, \bibinfo {author} {\bibfnamefont {B.~K.}\ \bibnamefont {Breedlove}},\ and\ \bibinfo {author} {\bibfnamefont {M.}~\bibnamefont {Yamashita}},\ }\bibfield  {title} {\bibinfo {title} {Observation and electric current control of a local spin in a single-molecule magnet},\ }\href {https://doi.org/10.1038/ncomms1210} {\bibfield  {journal} {\bibinfo  {journal} {Nat. Commun.}\ }\textbf {\bibinfo {volume} {2}},\ \bibinfo {pages} {217} (\bibinfo {year} {2011})}\BibitemShut {NoStop}%
\bibitem [{\citenamefont {Mugarza}\ \emph {et~al.}(2011)\citenamefont {Mugarza}, \citenamefont {Krull}, \citenamefont {Robles}, \citenamefont {Stepanow}, \citenamefont {Ceballos},\ and\ \citenamefont {Gambardella}}]{Mugarza2011}%
  \BibitemOpen
  \bibfield  {author} {\bibinfo {author} {\bibfnamefont {A.}~\bibnamefont {Mugarza}}, \bibinfo {author} {\bibfnamefont {C.}~\bibnamefont {Krull}}, \bibinfo {author} {\bibfnamefont {R.}~\bibnamefont {Robles}}, \bibinfo {author} {\bibfnamefont {S.}~\bibnamefont {Stepanow}}, \bibinfo {author} {\bibfnamefont {G.}~\bibnamefont {Ceballos}},\ and\ \bibinfo {author} {\bibfnamefont {P.}~\bibnamefont {Gambardella}},\ }\bibfield  {title} {\bibinfo {title} {Spin coupling and relaxation inside molecule--metal contacts},\ }\href {https://doi.org/10.1038/ncomms1497} {\bibfield  {journal} {\bibinfo  {journal} {Nat. Commun.}\ }\textbf {\bibinfo {volume} {2}},\ \bibinfo {pages} {490} (\bibinfo {year} {2011})}\BibitemShut {NoStop}%
\bibitem [{\citenamefont {Franke}\ \emph {et~al.}(2011)\citenamefont {Franke}, \citenamefont {Schulze},\ and\ \citenamefont {Pascual}}]{Franke2011}%
  \BibitemOpen
  \bibfield  {author} {\bibinfo {author} {\bibfnamefont {K.~J.}\ \bibnamefont {Franke}}, \bibinfo {author} {\bibfnamefont {G.}~\bibnamefont {Schulze}},\ and\ \bibinfo {author} {\bibfnamefont {J.~I.}\ \bibnamefont {Pascual}},\ }\bibfield  {title} {\bibinfo {title} {Competition of superconducting phenomena and {Kondo} screening at the nanoscale},\ }\href {https://doi.org/10.1126/science.1202204} {\bibfield  {journal} {\bibinfo  {journal} {Science}\ }\textbf {\bibinfo {volume} {332}},\ \bibinfo {pages} {940} (\bibinfo {year} {2011})}\BibitemShut {NoStop}%
\bibitem [{\citenamefont {Minamitani}\ \emph {et~al.}(2012)\citenamefont {Minamitani}, \citenamefont {Tsukahara}, \citenamefont {Matsunaka}, \citenamefont {Kim}, \citenamefont {Takagi},\ and\ \citenamefont {Kawai}}]{Minamitani2012}%
  \BibitemOpen
  \bibfield  {author} {\bibinfo {author} {\bibfnamefont {E.}~\bibnamefont {Minamitani}}, \bibinfo {author} {\bibfnamefont {N.}~\bibnamefont {Tsukahara}}, \bibinfo {author} {\bibfnamefont {D.}~\bibnamefont {Matsunaka}}, \bibinfo {author} {\bibfnamefont {Y.}~\bibnamefont {Kim}}, \bibinfo {author} {\bibfnamefont {N.}~\bibnamefont {Takagi}},\ and\ \bibinfo {author} {\bibfnamefont {M.}~\bibnamefont {Kawai}},\ }\bibfield  {title} {\bibinfo {title} {Symmetry-driven novel {Kondo} effect in a molecule},\ }\href {https://doi.org/10.1103/PhysRevLett.109.086602} {\bibfield  {journal} {\bibinfo  {journal} {Phys. Rev. Lett.}\ }\textbf {\bibinfo {volume} {109}},\ \bibinfo {pages} {086602} (\bibinfo {year} {2012})}\BibitemShut {NoStop}%
\bibitem [{\citenamefont {Zhang}\ \emph {et~al.}(2013)\citenamefont {Zhang}, \citenamefont {Kahle}, \citenamefont {Herden}, \citenamefont {Stroh}, \citenamefont {Mayor}, \citenamefont {Schlickum}, \citenamefont {Ternes}, \citenamefont {Wahl},\ and\ \citenamefont {Kern}}]{Zhang2013}%
  \BibitemOpen
  \bibfield  {author} {\bibinfo {author} {\bibfnamefont {Y.-h.}\ \bibnamefont {Zhang}}, \bibinfo {author} {\bibfnamefont {S.}~\bibnamefont {Kahle}}, \bibinfo {author} {\bibfnamefont {T.}~\bibnamefont {Herden}}, \bibinfo {author} {\bibfnamefont {C.}~\bibnamefont {Stroh}}, \bibinfo {author} {\bibfnamefont {M.}~\bibnamefont {Mayor}}, \bibinfo {author} {\bibfnamefont {U.}~\bibnamefont {Schlickum}}, \bibinfo {author} {\bibfnamefont {M.}~\bibnamefont {Ternes}}, \bibinfo {author} {\bibfnamefont {P.}~\bibnamefont {Wahl}},\ and\ \bibinfo {author} {\bibfnamefont {K.}~\bibnamefont {Kern}},\ }\bibfield  {title} {\bibinfo {title} {Temperature and magnetic field dependence of a kondo system in the weak coupling regime},\ }\href {https://doi.org/10.1038/ncomms3110} {\bibfield  {journal} {\bibinfo  {journal} {Nat. Commun.}\ }\textbf {\bibinfo {volume} {4}},\ \bibinfo {pages} {2110} (\bibinfo {year} {2013})}\BibitemShut {NoStop}%
\bibitem [{\citenamefont {Verlhac}\ \emph {et~al.}(2019)\citenamefont {Verlhac}, \citenamefont {Bachellier}, \citenamefont {Garnier}, \citenamefont {Ormaza}, \citenamefont {Abufager}, \citenamefont {Robles}, \citenamefont {M.-L.}, \citenamefont {Ternes}, \citenamefont {Lorente},\ and\ \citenamefont {Limot}}]{Verlhac2019}%
  \BibitemOpen
  \bibfield  {author} {\bibinfo {author} {\bibfnamefont {B.}~\bibnamefont {Verlhac}}, \bibinfo {author} {\bibfnamefont {N.}~\bibnamefont {Bachellier}}, \bibinfo {author} {\bibfnamefont {L.}~\bibnamefont {Garnier}}, \bibinfo {author} {\bibfnamefont {M.}~\bibnamefont {Ormaza}}, \bibinfo {author} {\bibfnamefont {P.}~\bibnamefont {Abufager}}, \bibinfo {author} {\bibfnamefont {R.}~\bibnamefont {Robles}}, \bibinfo {author} {\bibfnamefont {B.}~\bibnamefont {M.-L.}}, \bibinfo {author} {\bibfnamefont {M.}~\bibnamefont {Ternes}}, \bibinfo {author} {\bibfnamefont {N.}~\bibnamefont {Lorente}},\ and\ \bibinfo {author} {\bibfnamefont {L.}~\bibnamefont {Limot}},\ }\bibfield  {title} {\bibinfo {title} {Atomic-scale spin sensing with a single molecule at the apex of a scanning tunneling microscope},\ }\href {https://doi.org/10.1126/science.aax8222} {\bibfield  {journal} {\bibinfo  {journal} {Science}\ }\textbf {\bibinfo {volume} {366}},\ \bibinfo {pages} {623} (\bibinfo {year} {2019})}\BibitemShut {NoStop}%
\bibitem [{\citenamefont {Friedrich}\ \emph {et~al.}(2024)\citenamefont {Friedrich}, \citenamefont {Odobesko}, \citenamefont {Bouaziz}, \citenamefont {Lounis},\ and\ \citenamefont {Bode}}]{Friedrich2024}%
  \BibitemOpen
  \bibfield  {author} {\bibinfo {author} {\bibfnamefont {F.}~\bibnamefont {Friedrich}}, \bibinfo {author} {\bibfnamefont {A.}~\bibnamefont {Odobesko}}, \bibinfo {author} {\bibfnamefont {J.}~\bibnamefont {Bouaziz}}, \bibinfo {author} {\bibfnamefont {S.}~\bibnamefont {Lounis}},\ and\ \bibinfo {author} {\bibfnamefont {M.}~\bibnamefont {Bode}},\ }\bibfield  {title} {\bibinfo {title} {Evidence for spinarons in {Co} adatoms},\ }\href {https://doi.org/10.1038/s41567-023-02262-6} {\bibfield  {journal} {\bibinfo  {journal} {Nat. Phys.}\ }\textbf {\bibinfo {volume} {20}},\ \bibinfo {pages} {28} (\bibinfo {year} {2024})}\BibitemShut {NoStop}%
\bibitem [{\citenamefont {Natterer}\ \emph {et~al.}(2017)\citenamefont {Natterer}, \citenamefont {Yang}, \citenamefont {Paul}, \citenamefont {Willke}, \citenamefont {Choi}, \citenamefont {Greber}, \citenamefont {Heinrich},\ and\ \citenamefont {Lutz}}]{Natterer2017}%
  \BibitemOpen
  \bibfield  {author} {\bibinfo {author} {\bibfnamefont {F.~D.}\ \bibnamefont {Natterer}}, \bibinfo {author} {\bibfnamefont {K.}~\bibnamefont {Yang}}, \bibinfo {author} {\bibfnamefont {W.}~\bibnamefont {Paul}}, \bibinfo {author} {\bibfnamefont {P.}~\bibnamefont {Willke}}, \bibinfo {author} {\bibfnamefont {T.}~\bibnamefont {Choi}}, \bibinfo {author} {\bibfnamefont {T.}~\bibnamefont {Greber}}, \bibinfo {author} {\bibfnamefont {A.~J.}\ \bibnamefont {Heinrich}},\ and\ \bibinfo {author} {\bibfnamefont {C.~P.}\ \bibnamefont {Lutz}},\ }\bibfield  {title} {\bibinfo {title} {Reading and writing single-atom magnets},\ }\href {https://doi.org/10.1038/nature21371} {\bibfield  {journal} {\bibinfo  {journal} {Nature}\ }\textbf {\bibinfo {volume} {543}},\ \bibinfo {pages} {226} (\bibinfo {year} {2017})}\BibitemShut {NoStop}%
\bibitem [{\citenamefont {Trishin}\ \emph {et~al.}(2023)\citenamefont {Trishin}, \citenamefont {Lotze}, \citenamefont {Lohss}, \citenamefont {Franceschi}, \citenamefont {Glazman}, \citenamefont {von Oppen},\ and\ \citenamefont {Franke}}]{Trishi2023}%
  \BibitemOpen
  \bibfield  {author} {\bibinfo {author} {\bibfnamefont {S.}~\bibnamefont {Trishin}}, \bibinfo {author} {\bibfnamefont {C.}~\bibnamefont {Lotze}}, \bibinfo {author} {\bibfnamefont {F.}~\bibnamefont {Lohss}}, \bibinfo {author} {\bibfnamefont {G.}~\bibnamefont {Franceschi}}, \bibinfo {author} {\bibfnamefont {L.~I.}\ \bibnamefont {Glazman}}, \bibinfo {author} {\bibfnamefont {F.}~\bibnamefont {von Oppen}},\ and\ \bibinfo {author} {\bibfnamefont {K.~J.}\ \bibnamefont {Franke}},\ }\bibfield  {title} {\bibinfo {title} {Tuning a two-impurity {Kondo} system by a {M}oir\'{e} superstructure},\ }\href {https://doi.org/10.1103/PhysRevLett.130.176201} {\bibfield  {journal} {\bibinfo  {journal} {Phys. Rev. Lett.}\ }\textbf {\bibinfo {volume} {130}},\ \bibinfo {pages} {176201} (\bibinfo {year} {2023})}\BibitemShut {NoStop}%
\bibitem [{\citenamefont {Hirjibehedin}\ \emph {et~al.}(2006)\citenamefont {Hirjibehedin}, \citenamefont {Lutz},\ and\ \citenamefont {Heinrich}}]{Hirjibehedin2006}%
  \BibitemOpen
  \bibfield  {author} {\bibinfo {author} {\bibfnamefont {C.~F.}\ \bibnamefont {Hirjibehedin}}, \bibinfo {author} {\bibfnamefont {C.~P.}\ \bibnamefont {Lutz}},\ and\ \bibinfo {author} {\bibfnamefont {A.~J.}\ \bibnamefont {Heinrich}},\ }\bibfield  {title} {\bibinfo {title} {Spin coupling in engineered atomic structures},\ }\href {https://doi.org/10.1126/science.1125398} {\bibfield  {journal} {\bibinfo  {journal} {Science}\ }\textbf {\bibinfo {volume} {312}},\ \bibinfo {pages} {1021} (\bibinfo {year} {2006})}\BibitemShut {NoStop}%
\bibitem [{\citenamefont {Bork}\ \emph {et~al.}(2011)\citenamefont {Bork}, \citenamefont {Zhang}, \citenamefont {Diekh{\"o}ner}, \citenamefont {Borda}, \citenamefont {Simon}, \citenamefont {Kroha}, \citenamefont {Wahl},\ and\ \citenamefont {Kern}}]{Bork2011}%
  \BibitemOpen
  \bibfield  {author} {\bibinfo {author} {\bibfnamefont {J.}~\bibnamefont {Bork}}, \bibinfo {author} {\bibfnamefont {Y.}~\bibnamefont {Zhang}}, \bibinfo {author} {\bibfnamefont {L.}~\bibnamefont {Diekh{\"o}ner}}, \bibinfo {author} {\bibfnamefont {L.}~\bibnamefont {Borda}}, \bibinfo {author} {\bibfnamefont {P.}~\bibnamefont {Simon}}, \bibinfo {author} {\bibfnamefont {J.}~\bibnamefont {Kroha}}, \bibinfo {author} {\bibfnamefont {P.}~\bibnamefont {Wahl}},\ and\ \bibinfo {author} {\bibfnamefont {K.}~\bibnamefont {Kern}},\ }\bibfield  {title} {\bibinfo {title} {A tunable two-impurity {K}ondo system in an atomic point contact},\ }\href {https://doi.org/10.1038/nphys2076} {\bibfield  {journal} {\bibinfo  {journal} {Nat. Phys.}\ }\textbf {\bibinfo {volume} {7}},\ \bibinfo {pages} {901} (\bibinfo {year} {2011})}\BibitemShut {NoStop}%
\bibitem [{\citenamefont {Spinelli}\ \emph {et~al.}(2015)\citenamefont {Spinelli}, \citenamefont {Gerrits}, \citenamefont {Toskovic}, \citenamefont {Bryant}, \citenamefont {Ternes},\ and\ \citenamefont {Otte}}]{Spinelli2015}%
  \BibitemOpen
  \bibfield  {author} {\bibinfo {author} {\bibfnamefont {A.}~\bibnamefont {Spinelli}}, \bibinfo {author} {\bibfnamefont {M.}~\bibnamefont {Gerrits}}, \bibinfo {author} {\bibfnamefont {R.}~\bibnamefont {Toskovic}}, \bibinfo {author} {\bibfnamefont {B.}~\bibnamefont {Bryant}}, \bibinfo {author} {\bibfnamefont {M.}~\bibnamefont {Ternes}},\ and\ \bibinfo {author} {\bibfnamefont {A.~F.}\ \bibnamefont {Otte}},\ }\bibfield  {title} {\bibinfo {title} {Exploring the phase diagram of the two-impurity kondo problem},\ }\href {https://doi.org/10.1038/ncomms10046} {\bibfield  {journal} {\bibinfo  {journal} {Nat. Commun.}\ }\textbf {\bibinfo {volume} {6}},\ \bibinfo {pages} {10046} (\bibinfo {year} {2015})}\BibitemShut {NoStop}%
\bibitem [{\citenamefont {Ternes}(2017)}]{Ternes2017}%
  \BibitemOpen
  \bibfield  {author} {\bibinfo {author} {\bibfnamefont {M.}~\bibnamefont {Ternes}},\ }\bibfield  {title} {\bibinfo {title} {Probing magnetic excitations and correlations in single and coupled spin systems with scanning tunneling spectroscopy},\ }\href {https://doi.org/10.1016/j.progsurf.2017.01.001} {\bibfield  {journal} {\bibinfo  {journal} {Prog. Surf. Sci.}\ }\textbf {\bibinfo {volume} {92}},\ \bibinfo {pages} {83} (\bibinfo {year} {2017})}\BibitemShut {NoStop}%
\bibitem [{\citenamefont {Choi}\ \emph {et~al.}(2019)\citenamefont {Choi}, \citenamefont {Lorente}, \citenamefont {Wiebe}, \citenamefont {von Bergmann}, \citenamefont {Otte},\ and\ \citenamefont {Heinrich}}]{Choi2019}%
  \BibitemOpen
  \bibfield  {author} {\bibinfo {author} {\bibfnamefont {D.-J.}\ \bibnamefont {Choi}}, \bibinfo {author} {\bibfnamefont {N.}~\bibnamefont {Lorente}}, \bibinfo {author} {\bibfnamefont {J.}~\bibnamefont {Wiebe}}, \bibinfo {author} {\bibfnamefont {K.}~\bibnamefont {von Bergmann}}, \bibinfo {author} {\bibfnamefont {A.~F.}\ \bibnamefont {Otte}},\ and\ \bibinfo {author} {\bibfnamefont {A.~J.}\ \bibnamefont {Heinrich}},\ }\bibfield  {title} {\bibinfo {title} {Colloquium: Atomic spin chains on surfaces},\ }\href {https://doi.org/10.1103/RevModPhys.91.041001} {\bibfield  {journal} {\bibinfo  {journal} {Rev. Mod. Phys.}\ }\textbf {\bibinfo {volume} {91}},\ \bibinfo {pages} {041001} (\bibinfo {year} {2019})}\BibitemShut {NoStop}%
\bibitem [{\citenamefont {Loth}\ \emph {et~al.}(2012)\citenamefont {Loth}, \citenamefont {Baumann}, \citenamefont {Lutz}, \citenamefont {Eigler},\ and\ \citenamefont {Heinrich}}]{Loth2012}%
  \BibitemOpen
  \bibfield  {author} {\bibinfo {author} {\bibfnamefont {S.}~\bibnamefont {Loth}}, \bibinfo {author} {\bibfnamefont {S.}~\bibnamefont {Baumann}}, \bibinfo {author} {\bibfnamefont {C.~P.}\ \bibnamefont {Lutz}}, \bibinfo {author} {\bibfnamefont {D.~M.}\ \bibnamefont {Eigler}},\ and\ \bibinfo {author} {\bibfnamefont {A.~J.}\ \bibnamefont {Heinrich}},\ }\bibfield  {title} {\bibinfo {title} {Bistability in atomic-scale antiferromagnets},\ }\href {https://doi.org/10.1126/science.1214131} {\bibfield  {journal} {\bibinfo  {journal} {Science}\ }\textbf {\bibinfo {volume} {335}},\ \bibinfo {pages} {196} (\bibinfo {year} {2012})}\BibitemShut {NoStop}%
\bibitem [{\citenamefont {Ako}\ \emph {et~al.}(2011)\citenamefont {Ako}, \citenamefont {Jens}, \citenamefont {Bruno},\ and\ \citenamefont {Roland}}]{Ako2011}%
  \BibitemOpen
  \bibfield  {author} {\bibinfo {author} {\bibfnamefont {K.~A.}\ \bibnamefont {Ako}}, \bibinfo {author} {\bibfnamefont {W.}~\bibnamefont {Jens}}, \bibinfo {author} {\bibfnamefont {C.}~\bibnamefont {Bruno}},\ and\ \bibinfo {author} {\bibfnamefont {W.}~\bibnamefont {Roland}},\ }\bibfield  {title} {\bibinfo {title} {Realizing all-spin-based logic operations atom by atom},\ }\href {https://doi.org/10.1126/science.1201725} {\bibfield  {journal} {\bibinfo  {journal} {Science}\ }\textbf {\bibinfo {volume} {332}},\ \bibinfo {pages} {1062} (\bibinfo {year} {2011})}\BibitemShut {NoStop}%
\bibitem [{\citenamefont {Wang}\ \emph {et~al.}(2022)\citenamefont {Wang}, \citenamefont {Berdonces-Layunta}, \citenamefont {Friedrich}, \citenamefont {Vilas-Varela}, \citenamefont {Patrick~Calupitan}, \citenamefont {Ignacio~Pascual}, \citenamefont {Pena}, \citenamefont {Casanova}, \citenamefont {Corso},\ and\ \citenamefont {de~Oteyza}}]{WangDimas2022}%
  \BibitemOpen
  \bibfield  {author} {\bibinfo {author} {\bibfnamefont {T.}~\bibnamefont {Wang}}, \bibinfo {author} {\bibfnamefont {A.}~\bibnamefont {Berdonces-Layunta}}, \bibinfo {author} {\bibfnamefont {N.}~\bibnamefont {Friedrich}}, \bibinfo {author} {\bibfnamefont {M.}~\bibnamefont {Vilas-Varela}}, \bibinfo {author} {\bibfnamefont {J.}~\bibnamefont {Patrick~Calupitan}}, \bibinfo {author} {\bibfnamefont {J.}~\bibnamefont {Ignacio~Pascual}}, \bibinfo {author} {\bibfnamefont {D.}~\bibnamefont {Pena}}, \bibinfo {author} {\bibfnamefont {D.}~\bibnamefont {Casanova}}, \bibinfo {author} {\bibfnamefont {M.}~\bibnamefont {Corso}},\ and\ \bibinfo {author} {\bibfnamefont {D.~G.}\ \bibnamefont {de~Oteyza}},\ }\bibfield  {title} {\bibinfo {title} {Aza-triangulene: On-surface synthesis and electronic and magnetic properties},\ }\href {https://doi.org/10.1021/jacs.1c12618} {\bibfield  {journal} {\bibinfo  {journal} {J. Am. Chem. Soc.}\ }\textbf {\bibinfo {volume} {144}},\ \bibinfo {pages} {4522} (\bibinfo {year} {2022})}\BibitemShut
  {NoStop}%
\bibitem [{\citenamefont {Mishra}\ \emph {et~al.}(2020)\citenamefont {Mishra}, \citenamefont {Beyer}, \citenamefont {Eimre}, \citenamefont {Kezilebieke}, \citenamefont {Berger}, \citenamefont {Gröning}, \citenamefont {Pignedoli}, \citenamefont {Müllen}, \citenamefont {Liljeroth}, \citenamefont {Ruffieux}, \citenamefont {Feng},\ and\ \citenamefont {Fasel}}]{Mishra2020}%
  \BibitemOpen
  \bibfield  {author} {\bibinfo {author} {\bibfnamefont {S.}~\bibnamefont {Mishra}}, \bibinfo {author} {\bibfnamefont {D.}~\bibnamefont {Beyer}}, \bibinfo {author} {\bibfnamefont {K.}~\bibnamefont {Eimre}}, \bibinfo {author} {\bibfnamefont {S.}~\bibnamefont {Kezilebieke}}, \bibinfo {author} {\bibfnamefont {R.}~\bibnamefont {Berger}}, \bibinfo {author} {\bibfnamefont {O.}~\bibnamefont {Gröning}}, \bibinfo {author} {\bibfnamefont {C.~A.}\ \bibnamefont {Pignedoli}}, \bibinfo {author} {\bibfnamefont {K.}~\bibnamefont {Müllen}}, \bibinfo {author} {\bibfnamefont {P.}~\bibnamefont {Liljeroth}}, \bibinfo {author} {\bibfnamefont {P.}~\bibnamefont {Ruffieux}}, \bibinfo {author} {\bibfnamefont {X.}~\bibnamefont {Feng}},\ and\ \bibinfo {author} {\bibfnamefont {R.}~\bibnamefont {Fasel}},\ }\bibfield  {title} {\bibinfo {title} {Topological frustration induces unconventional magnetism in a nanographene},\ }\href {https://doi.org/10.1038/s41565-019-0577-9} {\bibfield  {journal} {\bibinfo  {journal} {Nat. Nanotechnol.}\
  }\textbf {\bibinfo {volume} {15}},\ \bibinfo {pages} {22} (\bibinfo {year} {2020})}\BibitemShut {NoStop}%
\bibitem [{\citenamefont {Song}\ \emph {et~al.}(2024)\citenamefont {Song}, \citenamefont {Sole}, \citenamefont {Matej}, \citenamefont {Li}, \citenamefont {Stetsovych}, \citenamefont {Soler}, \citenamefont {Yang}, \citenamefont {Telychko}, \citenamefont {Li}, \citenamefont {Kumar}, \citenamefont {Chen}, \citenamefont {Edalatmanesh}, \citenamefont {Brabec}, \citenamefont {Veis}, \citenamefont {Wu}, \citenamefont {Jelinek},\ and\ \citenamefont {Lu}}]{SongLu2024}%
  \BibitemOpen
  \bibfield  {author} {\bibinfo {author} {\bibfnamefont {S.}~\bibnamefont {Song}}, \bibinfo {author} {\bibfnamefont {A.~P.}\ \bibnamefont {Sole}}, \bibinfo {author} {\bibfnamefont {A.}~\bibnamefont {Matej}}, \bibinfo {author} {\bibfnamefont {G.}~\bibnamefont {Li}}, \bibinfo {author} {\bibfnamefont {O.}~\bibnamefont {Stetsovych}}, \bibinfo {author} {\bibfnamefont {D.}~\bibnamefont {Soler}}, \bibinfo {author} {\bibfnamefont {H.}~\bibnamefont {Yang}}, \bibinfo {author} {\bibfnamefont {M.}~\bibnamefont {Telychko}}, \bibinfo {author} {\bibfnamefont {J.}~\bibnamefont {Li}}, \bibinfo {author} {\bibfnamefont {M.}~\bibnamefont {Kumar}}, \bibinfo {author} {\bibfnamefont {Q.}~\bibnamefont {Chen}}, \bibinfo {author} {\bibfnamefont {S.}~\bibnamefont {Edalatmanesh}}, \bibinfo {author} {\bibfnamefont {J.}~\bibnamefont {Brabec}}, \bibinfo {author} {\bibfnamefont {L.}~\bibnamefont {Veis}}, \bibinfo {author} {\bibfnamefont {J.}~\bibnamefont {Wu}}, \bibinfo {author} {\bibfnamefont {P.}~\bibnamefont {Jelinek}},\ and\ \bibinfo
  {author} {\bibfnamefont {J.}~\bibnamefont {Lu}},\ }\bibfield  {title} {\bibinfo {title} {Highly entangled polyradical nanographene with coexisting strong correlation and topological frustration},\ }\bibfield  {journal} {\bibinfo  {journal} {Nat. Chem.}\ }\textbf {\bibinfo {volume} {16}},\ \href {https://doi.org/10.1038/s41557-024-01453-9} {10.1038/s41557-024-01453-9} (\bibinfo {year} {2024})\BibitemShut {NoStop}%
\bibitem [{\citenamefont {Sun}\ \emph {et~al.}(2022)\citenamefont {Sun}, \citenamefont {Mateo}, \citenamefont {Robles}, \citenamefont {Ruffieux}, \citenamefont {Bottari}, \citenamefont {Torres}, \citenamefont {Fasel},\ and\ \citenamefont {Lorente}}]{SunLorente2022}%
  \BibitemOpen
  \bibfield  {author} {\bibinfo {author} {\bibfnamefont {Q.}~\bibnamefont {Sun}}, \bibinfo {author} {\bibfnamefont {L.~M.}\ \bibnamefont {Mateo}}, \bibinfo {author} {\bibfnamefont {R.}~\bibnamefont {Robles}}, \bibinfo {author} {\bibfnamefont {P.}~\bibnamefont {Ruffieux}}, \bibinfo {author} {\bibfnamefont {G.}~\bibnamefont {Bottari}}, \bibinfo {author} {\bibfnamefont {T.}~\bibnamefont {Torres}}, \bibinfo {author} {\bibfnamefont {R.}~\bibnamefont {Fasel}},\ and\ \bibinfo {author} {\bibfnamefont {N.}~\bibnamefont {Lorente}},\ }\bibfield  {title} {\bibinfo {title} {Magnetic interplay between $\pi$-electrons of open-shell porphyrins and $d$-electrons of their central transition metal ions},\ }\bibfield  {journal} {\bibinfo  {journal} {Adv. Sci.}\ }\textbf {\bibinfo {volume} {9}},\ \href {https://doi.org/10.1002/advs.202105906} {10.1002/advs.202105906} (\bibinfo {year} {2022})\BibitemShut {NoStop}%
\bibitem [{\citenamefont {Zhao}\ \emph {et~al.}(2023)\citenamefont {Zhao}, \citenamefont {Jiang}, \citenamefont {Li}, \citenamefont {Liu}, \citenamefont {Zhu}, \citenamefont {Pizzochero}, \citenamefont {Kaxiras}, \citenamefont {Guan}, \citenamefont {Li}, \citenamefont {Zheng}, \citenamefont {Liu}, \citenamefont {Jia}, \citenamefont {Qin}, \citenamefont {Zhuang},\ and\ \citenamefont {Wang}}]{ZhaoWang2023}%
  \BibitemOpen
  \bibfield  {author} {\bibinfo {author} {\bibfnamefont {Y.}~\bibnamefont {Zhao}}, \bibinfo {author} {\bibfnamefont {K.}~\bibnamefont {Jiang}}, \bibinfo {author} {\bibfnamefont {C.}~\bibnamefont {Li}}, \bibinfo {author} {\bibfnamefont {Y.}~\bibnamefont {Liu}}, \bibinfo {author} {\bibfnamefont {G.}~\bibnamefont {Zhu}}, \bibinfo {author} {\bibfnamefont {M.}~\bibnamefont {Pizzochero}}, \bibinfo {author} {\bibfnamefont {E.}~\bibnamefont {Kaxiras}}, \bibinfo {author} {\bibfnamefont {D.}~\bibnamefont {Guan}}, \bibinfo {author} {\bibfnamefont {Y.}~\bibnamefont {Li}}, \bibinfo {author} {\bibfnamefont {H.}~\bibnamefont {Zheng}}, \bibinfo {author} {\bibfnamefont {C.}~\bibnamefont {Liu}}, \bibinfo {author} {\bibfnamefont {J.}~\bibnamefont {Jia}}, \bibinfo {author} {\bibfnamefont {M.}~\bibnamefont {Qin}}, \bibinfo {author} {\bibfnamefont {X.}~\bibnamefont {Zhuang}},\ and\ \bibinfo {author} {\bibfnamefont {S.}~\bibnamefont {Wang}},\ }\bibfield  {title} {\bibinfo {title} {Quantum nanomagnets in on-surface metal-free
  porphyrin chains},\ }\href {https://doi.org/10.1038/s41557-022-01061-5} {\bibfield  {journal} {\bibinfo  {journal} {Nat. Chem.}\ }\textbf {\bibinfo {volume} {15}},\ \bibinfo {pages} {53} (\bibinfo {year} {2023})}\BibitemShut {NoStop}%
\bibitem [{\citenamefont {He}\ \emph {et~al.}(2022)\citenamefont {He}, \citenamefont {Li}, \citenamefont {Castelli}, \citenamefont {Li}, \citenamefont {Zhang}, \citenamefont {Zhang}, \citenamefont {Li}, \citenamefont {Wang}, \citenamefont {Gao}, \citenamefont {Peng}, \citenamefont {Hou}, \citenamefont {Shen}, \citenamefont {L\"u}, \citenamefont {Wu}, \citenamefont {Hedeg\aa{}rd},\ and\ \citenamefont {Wang}}]{HeWang2022}%
  \BibitemOpen
  \bibfield  {author} {\bibinfo {author} {\bibfnamefont {Y.}~\bibnamefont {He}}, \bibinfo {author} {\bibfnamefont {N.}~\bibnamefont {Li}}, \bibinfo {author} {\bibfnamefont {I.~E.}\ \bibnamefont {Castelli}}, \bibinfo {author} {\bibfnamefont {R.}~\bibnamefont {Li}}, \bibinfo {author} {\bibfnamefont {Y.}~\bibnamefont {Zhang}}, \bibinfo {author} {\bibfnamefont {X.}~\bibnamefont {Zhang}}, \bibinfo {author} {\bibfnamefont {C.}~\bibnamefont {Li}}, \bibinfo {author} {\bibfnamefont {B.}~\bibnamefont {Wang}}, \bibinfo {author} {\bibfnamefont {S.}~\bibnamefont {Gao}}, \bibinfo {author} {\bibfnamefont {L.}~\bibnamefont {Peng}}, \bibinfo {author} {\bibfnamefont {S.}~\bibnamefont {Hou}}, \bibinfo {author} {\bibfnamefont {Z.}~\bibnamefont {Shen}}, \bibinfo {author} {\bibfnamefont {J.-T.}\ \bibnamefont {L\"u}}, \bibinfo {author} {\bibfnamefont {K.}~\bibnamefont {Wu}}, \bibinfo {author} {\bibfnamefont {P.}~\bibnamefont {Hedeg\aa{}rd}},\ and\ \bibinfo {author} {\bibfnamefont {Y.}~\bibnamefont {Wang}},\ }\bibfield  {title}
  {\bibinfo {title} {Observation of biradical spin coupling through hydrogen bonds},\ }\href {https://doi.org/10.1103/PhysRevLett.128.236401} {\bibfield  {journal} {\bibinfo  {journal} {Phys. Rev. Lett.}\ }\textbf {\bibinfo {volume} {128}},\ \bibinfo {pages} {236401} (\bibinfo {year} {2022})}\BibitemShut {NoStop}%
\bibitem [{\citenamefont {Langner}\ \emph {et~al.}(2007)\citenamefont {Langner}, \citenamefont {Tait}, \citenamefont {Lin}, \citenamefont {Rajadurai}, \citenamefont {Ruben},\ and\ \citenamefont {Kern}}]{Langner2007self}%
  \BibitemOpen
  \bibfield  {author} {\bibinfo {author} {\bibfnamefont {A.}~\bibnamefont {Langner}}, \bibinfo {author} {\bibfnamefont {S.~L.}\ \bibnamefont {Tait}}, \bibinfo {author} {\bibfnamefont {N.}~\bibnamefont {Lin}}, \bibinfo {author} {\bibfnamefont {C.}~\bibnamefont {Rajadurai}}, \bibinfo {author} {\bibfnamefont {M.}~\bibnamefont {Ruben}},\ and\ \bibinfo {author} {\bibfnamefont {K.}~\bibnamefont {Kern}},\ }\bibfield  {title} {\bibinfo {title} {Self-recognition and self-selection in multicomponent supramolecular coordination networks on surfaces},\ }\href {https://doi.org/10.1073/pnas.0704882104} {\bibfield  {journal} {\bibinfo  {journal} {Proc. Nat. Acad. Sci.}\ }\textbf {\bibinfo {volume} {104}},\ \bibinfo {pages} {17927} (\bibinfo {year} {2007})}\BibitemShut {NoStop}%
\bibitem [{\citenamefont {Tait}\ \emph {et~al.}(2008)\citenamefont {Tait}, \citenamefont {Wang}, \citenamefont {Costantini}, \citenamefont {Lin}, \citenamefont {Baraldi}, \citenamefont {Esch}, \citenamefont {Petaccia}, \citenamefont {Lizzit},\ and\ \citenamefont {Kern}}]{tait2008metal}%
  \BibitemOpen
  \bibfield  {author} {\bibinfo {author} {\bibfnamefont {S.~L.}\ \bibnamefont {Tait}}, \bibinfo {author} {\bibfnamefont {Y.}~\bibnamefont {Wang}}, \bibinfo {author} {\bibfnamefont {G.}~\bibnamefont {Costantini}}, \bibinfo {author} {\bibfnamefont {N.}~\bibnamefont {Lin}}, \bibinfo {author} {\bibfnamefont {A.}~\bibnamefont {Baraldi}}, \bibinfo {author} {\bibfnamefont {F.}~\bibnamefont {Esch}}, \bibinfo {author} {\bibfnamefont {L.}~\bibnamefont {Petaccia}}, \bibinfo {author} {\bibfnamefont {S.}~\bibnamefont {Lizzit}},\ and\ \bibinfo {author} {\bibfnamefont {K.}~\bibnamefont {Kern}},\ }\bibfield  {title} {\bibinfo {title} {{Metal-organic coordination interactions in Fe-Terephthalic acid networks on Cu (100)}},\ }\href {https://doi.org/10.1021/ja0778186} {\bibfield  {journal} {\bibinfo  {journal} {J. Am. Chem. Soc.}\ }\textbf {\bibinfo {volume} {130}},\ \bibinfo {pages} {2108} (\bibinfo {year} {2008})}\BibitemShut {NoStop}%
\bibitem [{\citenamefont {{\v{C}}echal}\ \emph {et~al.}(2014)\citenamefont {{\v{C}}echal}, \citenamefont {Kley}, \citenamefont {Kumagai}, \citenamefont {Schramm}, \citenamefont {Ruben}, \citenamefont {Stepanow},\ and\ \citenamefont {Kern}}]{vcechal2014convergent}%
  \BibitemOpen
  \bibfield  {author} {\bibinfo {author} {\bibfnamefont {J.}~\bibnamefont {{\v{C}}echal}}, \bibinfo {author} {\bibfnamefont {C.~S.}\ \bibnamefont {Kley}}, \bibinfo {author} {\bibfnamefont {T.}~\bibnamefont {Kumagai}}, \bibinfo {author} {\bibfnamefont {F.}~\bibnamefont {Schramm}}, \bibinfo {author} {\bibfnamefont {M.}~\bibnamefont {Ruben}}, \bibinfo {author} {\bibfnamefont {S.}~\bibnamefont {Stepanow}},\ and\ \bibinfo {author} {\bibfnamefont {K.}~\bibnamefont {Kern}},\ }\bibfield  {title} {\bibinfo {title} {Convergent and divergent two-dimensional coordination networks formed through substrate-activated or quenched alkynyl ligation},\ }\href {https://doi.org/10.1039/c4cc03723e} {\bibfield  {journal} {\bibinfo  {journal} {Chem. Commun.}\ }\textbf {\bibinfo {volume} {50}},\ \bibinfo {pages} {9973} (\bibinfo {year} {2014})}\BibitemShut {NoStop}%
\bibitem [{\citenamefont {Dong}\ \emph {et~al.}(2016)\citenamefont {Dong}, \citenamefont {Gao},\ and\ \citenamefont {Lin}}]{dong2016self}%
  \BibitemOpen
  \bibfield  {author} {\bibinfo {author} {\bibfnamefont {L.}~\bibnamefont {Dong}}, \bibinfo {author} {\bibfnamefont {Z.}~\bibnamefont {Gao}},\ and\ \bibinfo {author} {\bibfnamefont {N.}~\bibnamefont {Lin}},\ }\bibfield  {title} {\bibinfo {title} {Self-assembly of metal--organic coordination structures on surfaces},\ }\href {https://doi.org/j.progsurf.2016.08.001} {\bibfield  {journal} {\bibinfo  {journal} {Prog. Surf. Sci.}\ }\textbf {\bibinfo {volume} {91}},\ \bibinfo {pages} {101} (\bibinfo {year} {2016})}\BibitemShut {NoStop}%
\bibitem [{\citenamefont {Langner}\ \emph {et~al.}(2012)\citenamefont {Langner}, \citenamefont {Tait}, \citenamefont {Lin}, \citenamefont {Chandrasekar}, \citenamefont {Meded}, \citenamefont {Fink}, \citenamefont {Ruben},\ and\ \citenamefont {Kern}}]{langner2012selective}%
  \BibitemOpen
  \bibfield  {author} {\bibinfo {author} {\bibfnamefont {A.}~\bibnamefont {Langner}}, \bibinfo {author} {\bibfnamefont {S.~L.}\ \bibnamefont {Tait}}, \bibinfo {author} {\bibfnamefont {N.}~\bibnamefont {Lin}}, \bibinfo {author} {\bibfnamefont {R.}~\bibnamefont {Chandrasekar}}, \bibinfo {author} {\bibfnamefont {V.}~\bibnamefont {Meded}}, \bibinfo {author} {\bibfnamefont {K.}~\bibnamefont {Fink}}, \bibinfo {author} {\bibfnamefont {M.}~\bibnamefont {Ruben}},\ and\ \bibinfo {author} {\bibfnamefont {K.}~\bibnamefont {Kern}},\ }\bibfield  {title} {\bibinfo {title} {Selective coordination bonding in metallo-supramolecular systems on surfaces},\ }\href {https://doi.org/10.1002/anie.201108530} {\bibfield  {journal} {\bibinfo  {journal} {Angew. Chem. Int. Ed.}\ }\textbf {\bibinfo {volume} {51}},\ \bibinfo {pages} {4327} (\bibinfo {year} {2012})}\BibitemShut {NoStop}%
\bibitem [{\citenamefont {Karan}\ \emph {et~al.}(2016)\citenamefont {Karan}, \citenamefont {Li}, \citenamefont {Zhang}, \citenamefont {He}, \citenamefont {Hong}, \citenamefont {Song}, \citenamefont {L\"u}, \citenamefont {Wang}, \citenamefont {Peng}, \citenamefont {Wu}, \citenamefont {Michelitsch}, \citenamefont {Maurer}, \citenamefont {Diller}, \citenamefont {Reuter}, \citenamefont {Weismann},\ and\ \citenamefont {Berndt}}]{Karan2016}%
  \BibitemOpen
  \bibfield  {author} {\bibinfo {author} {\bibfnamefont {S.}~\bibnamefont {Karan}}, \bibinfo {author} {\bibfnamefont {N.}~\bibnamefont {Li}}, \bibinfo {author} {\bibfnamefont {Y.}~\bibnamefont {Zhang}}, \bibinfo {author} {\bibfnamefont {Y.}~\bibnamefont {He}}, \bibinfo {author} {\bibfnamefont {I.-P.}\ \bibnamefont {Hong}}, \bibinfo {author} {\bibfnamefont {H.}~\bibnamefont {Song}}, \bibinfo {author} {\bibfnamefont {J.-T.}\ \bibnamefont {L\"u}}, \bibinfo {author} {\bibfnamefont {Y.}~\bibnamefont {Wang}}, \bibinfo {author} {\bibfnamefont {L.}~\bibnamefont {Peng}}, \bibinfo {author} {\bibfnamefont {K.}~\bibnamefont {Wu}}, \bibinfo {author} {\bibfnamefont {G.~S.}\ \bibnamefont {Michelitsch}}, \bibinfo {author} {\bibfnamefont {R.~J.}\ \bibnamefont {Maurer}}, \bibinfo {author} {\bibfnamefont {K.}~\bibnamefont {Diller}}, \bibinfo {author} {\bibfnamefont {K.}~\bibnamefont {Reuter}}, \bibinfo {author} {\bibfnamefont {A.}~\bibnamefont {Weismann}},\ and\ \bibinfo {author} {\bibfnamefont {R.}~\bibnamefont {Berndt}},\
  }\bibfield  {title} {\bibinfo {title} {Spin manipulation by creation of single-molecule radical cations},\ }\href {https://doi.org/10.1103/PhysRevLett.116.027201} {\bibfield  {journal} {\bibinfo  {journal} {Phys. Rev. Lett.}\ }\textbf {\bibinfo {volume} {116}},\ \bibinfo {pages} {027201} (\bibinfo {year} {2016})}\BibitemShut {NoStop}%
\bibitem [{\citenamefont {Bocquet}\ \emph {et~al.}(2019)\citenamefont {Bocquet}, \citenamefont {Lorente}, \citenamefont {Berndt},\ and\ \citenamefont {Gruber}}]{Bocquet2019}%
  \BibitemOpen
  \bibfield  {author} {\bibinfo {author} {\bibfnamefont {M.-L.}\ \bibnamefont {Bocquet}}, \bibinfo {author} {\bibfnamefont {N.}~\bibnamefont {Lorente}}, \bibinfo {author} {\bibfnamefont {R.}~\bibnamefont {Berndt}},\ and\ \bibinfo {author} {\bibfnamefont {M.}~\bibnamefont {Gruber}},\ }\bibfield  {title} {\bibinfo {title} {Spin in a closed-shell organic molecule on a metal substrate generated by a sigmatropic reaction},\ }\href {https://doi.org/10.1002/anie.201812121} {\bibfield  {journal} {\bibinfo  {journal} {Angew. Chem. Int. Ed.}\ }\textbf {\bibinfo {volume} {131}},\ \bibinfo {pages} {831} (\bibinfo {year} {2019})}\BibitemShut {NoStop}%
\bibitem [{\citenamefont {Appelbaum}(1967)}]{Appelbaum1967}%
  \BibitemOpen
  \bibfield  {author} {\bibinfo {author} {\bibfnamefont {J.~A.}\ \bibnamefont {Appelbaum}},\ }\bibfield  {title} {\bibinfo {title} {Exchange model of zero-bias tunneling anomalies},\ }\href {https://doi.org/10.1103/PhysRev.154.633} {\bibfield  {journal} {\bibinfo  {journal} {Phys. Rev.}\ }\textbf {\bibinfo {volume} {154}},\ \bibinfo {pages} {633} (\bibinfo {year} {1967})}\BibitemShut {NoStop}%
\bibitem [{\citenamefont {Ternes}(2015)}]{Ternes2015}%
  \BibitemOpen
  \bibfield  {author} {\bibinfo {author} {\bibfnamefont {M.}~\bibnamefont {Ternes}},\ }\bibfield  {title} {\bibinfo {title} {Spin excitations and correlations in scanning tunneling spectroscopy},\ }\href {https://doi.org/10.1088/1367-2630/17/6/063016} {\bibfield  {journal} {\bibinfo  {journal} {New J. Phys.}\ }\textbf {\bibinfo {volume} {17}},\ \bibinfo {pages} {063016} (\bibinfo {year} {2015})}\BibitemShut {NoStop}%
\end{thebibliography}%

\end{document}